# Preliminary Guidelines for Electrode Positioning in Noninvasive Deep Brain Stimulation via Temporally Interfering Electric Fields

Mobina Zibandepour
*Faculty of Electrical Engineering*
*K. N. Toosi University of Technology*
Tehran, Iran
m.zibandepour@email.kntu.ac.ir

Akram Shojaei
*Faculty of Electrical Engineering*
*K. N. Toosi University of Technology*
Tehran, Iran
akram.shojaeibagheini@email.kntu.ac.ir

Arash Abbasi Larki
*Faculty of Electrical Engineering*
*K. N. Toosi University of Technology*
Tehran, Iran
arash.abbasilarki@email.kntu.ac.ir

Mehdi Delrobaei
*Department of Mechatronics*
*Faculty of Electrical Engineering*
*K. N. Toosi University of Technology*
Tehran, Iran
delrobaei@kntu.ac.ir

*Abstract*—Advancements in neurosurgical robotics have improved medical procedures, particularly deep brain stimulation, where robots combine human and machine intelligence to precisely implant electrodes in the brain. While effective, this procedure carries risks and side effects. Noninvasive deep brain stimulation (NIDBS) offers promise by making brain stimulation safer, more affordable, and accessible. However, NIDBS lacks guidelines for electrode placement. This study explores adapting robotic principles to enhance the accuracy of NIDBS targeting and provides preliminary guidelines for transcranial electrode placement. Safety is also emphasized, ensuring a balance between therapeutic effectiveness and patient safety by maintaining electric fields within safe limits.

*Index Terms*—Biomechatronic systems, transcranial stimulation, neuromodulation, electrode placement, current stimulation, electrode positioning, temporal interfering stimulation.

## I. INTRODUCTION

The use of robotics in neurosurgery has significantly advanced medical procedures like biopsies and deep brain stimulation (DBS), particularly in treating conditions like Parkinson's disease, obsessive-compulsive disorder, epilepsy, and depression. Robotic technology enables precise electrode implantation, improving treatment outcomes for many patients. However, it is important to note that implanting electrodes in the human brain remains a complex and potentially risky procedure. [1]. The success of a DBS procedure must be measured by electrode placement accuracy before determining clinical outcomes. Robots such as SurgiScope, NeuroMate, Renaissance, and ROSA are used in the operating room to improve the precision of asleep DBS. However, electrode placement for noninvasive DBS (NIDBS) is still a challenge, and the role of robotic concepts has yet to be fully explored [1], [2].To ensure safer and more accessible brain stimulation, researchers are exploring NIDBS techniques. However, the methods for achieving this goal remain a trial-and-error procedure for proper electrode positioning. Precision is a key factor, especially when targeting small structures deep in the brain. The implementation of robotic assistant concepts makes it possible to make electrode placement extremely precise in delivering brain stimulation treatments [2]. Noninvasive brain stimulation techniques such as tDCS and tACS use weak electric currents to regulate neural activity. A new technique called temporal interference (TI) stimulation is being developed to access deep brain regions, making the process easier and less risky [3]. Robotic techniques are emerging for noninvasive brain stimulation in transcranial electrode implantation, allowing for precise targeting and fixed tooling positions. Transcranial magnetic stimulation (TMS) is another noninvasive technique used to study cognition, emotion, sensation, and movement in normal subjects. However, TMS can also strongly stimulate surface regions, affecting multiple brain networks, and highlighting the need for in electrode positioning precision [4].The study evaluated the effectiveness of a robotic system for transcranial magnetic stimulation coil placement in treating depression. Researchers used finite element simulation and phantom measurements to determine functional relationships between target coordinates, current ratio, and electrode position. They proposed an appropriate approach for assessing electrode configurations in a realistic mouse model [5]. We used mathematical modeling to create a unique configuration of robotics, consisting of dual pairs of electrodes on a sphere's surface. The electric fields generated by these electrodes form the end effector. The spherical structure's radius was kept constant, and a reference coordinate system was established at the sphere's center. Each electrode can rotate along the *x*, *y*, and *z* axes, enabling precise control and measurement in three-dimensional space. The electrodes can navigate lateral, vertical and horizontal orientations, making them valuable tools in various medical interventions [4], [5]. These maneuvers enable vertical and horizontal tilting and orientation, making the system versatile and adaptable to various spatial tasks. With $n$ electrodes, each contributes two degrees of freedom, resulting in a collective degree of $2n$. With two pairs of electrodes, the system has 8 degrees of freedom, allowing for intricate spatial tasks and precise object orientation control, ensuring precision and flexibility [6]. The term "end-effector" refers to the specific stimulation region within a system, creating a virtual domain. The study emphasizes precise end-effector positioning but does not consider end-effector orientation. Multiple end effectors can occur in specific scenarios, emphasizing the importance of accurate location determination for optimal performance [7]. The robotic system with dual-electrode pairs on a sphere has holonomic constraints, enhancing precision and control. These constraints enable seamless coordination, efficiency in spatial tasks, optimal end-effector positioning, and adaptability to various spatial tasks, including noninvasive brain stimulation techniques like transcranial electrical and magnetic stimulation [8].

The paper is organized as follows: Section II refers to the prior research endeavor in this field. Section III provides a comprehensive description of the methodology employed in this study. The outcomes of the suggested research are outlined in Section IV, while an interpretation of our findings is provided in Section V. Finally, the concluding remarks are provided in Section VI.

## II. RELATED WORKS

Researchers have developed various methods to compute electric potential and field distributions during non-invasive deep brain stimulation (NIDBS) in brain models [9]. Two automatic algorithms were used for the homogeneous model and extracellular potentials for the microscopic approach. The results showed that the activated area was primarily situated at depth, and the optimization algorithms showed significant accuracy in estimating stimulation parameters. A novel technique called multipair temporal interference stimulation (TIS) was introduced, which involved using more than two electrode pairs to enhance the effectiveness of focused deep brain stimulation [10]. The study suggested employing ten electrode pairs for selective deep target stimulation without affecting neocortical regions. An innovative solution involving unsupervised neural networks (USNNs) was proposed to optimize high-definition electrodes for precise tTIS targeting [11]. Computational simulations on 16 realistic head models demonstrated the robustness of tTIS and reduced mis-stimulation compared to other methods. The USNN demonstrated efficient optimization of electrode currents for tTIS, surpassing conventional studies in terms of speed and accuracy. Despite minor disadvantages, the study emphasized the potential of interference stimulation in modulating the deep brain and holds promise for clinical testing and validation in brain stimulation using electric fields.

## III. METHODOLOGY

In the pursuit of simulating transcranial temporal interference brain stimulation (tTIs), a dual-pair electrode configuration [12] was employed. The computational framework was predicated on the quasi-static [13] finite element method (FEM), which was established by deriving the electrostatic Laplace equation

$$\nabla \cdot (\sigma \nabla V) = 0 \quad (1)$$

where $\sigma$ represents electrical conductivity, and $V$ denotes electrical potential. Besides, the relationship between electric field $E$ and electrical potential $V$ is given as:

$$E = -\nabla V \quad (2)$$

The current density [14], a component of Maxwell's equations, is also represented as

$$\mathbf{J} = \sigma \cdot \mathbf{E} \quad (3)$$

Hence, the current density can be rewritten as

$$\mathbf{J} = -\sigma \nabla V \quad (4)$$

In order to impose a continuity condition on the current density, we have

$$\nabla \cdot \mathbf{J} = \nabla \cdot (\sigma \nabla V) \quad (5)$$

The two electrode pairs yield electric fields [11] denoted as $\mathbf{E}_1$ and $\mathbf{E}_2$, representing the electric fields generated at the spatial location $\mathbf{r}_{(x,y,z)}$ with $\mathbf{n}$ serving as a unit vector direction. To determine the envelope amplitude of temporal interference, amplitude modulation is applied and expressed as follows:

$$|\mathbf{E}_{\text{TI-AM}}(\mathbf{n}\text{-}\mathbf{r})| = |\mathbf{E}_1(\mathbf{r}) + \mathbf{E}_2(\mathbf{r}) \cdot \mathbf{n}| - |\mathbf{E}_1(\mathbf{r}) - \mathbf{E}_2(\mathbf{r}) \cdot \mathbf{n}| \quad (6)$$

The amplitude $|\mathbf{E}_{\text{AM}}(\mathbf{r})|$ is defined as [3]:

$$|\mathbf{E}_{\text{AM}}(\mathbf{r})| = \begin{cases} 2|\mathbf{E}_2(\mathbf{r})| & \text{if } |\mathbf{E}_2(\mathbf{r})| < |\mathbf{E}_1(\mathbf{r})|\cos\gamma \\ 2\frac{\mathbf{E}_2(\mathbf{r})(\mathbf{E}_1(\mathbf{r}) - \mathbf{E}_2(\mathbf{r}))}{\mathbf{E}_1(\mathbf{r}) - \mathbf{E}_2(\mathbf{r})} & \text{otherwise} \end{cases} \quad (7)$$

To compute and analyze the maximum electric field in various directions, the angle $\gamma$ between $\mathbf{E}_1$ and $\mathbf{E}_2$ must be less than 45°. We investigated three scenarios and evaluated how three variables impact the stimulation pattern within the end effector's workspace Fig. 1.A. The placement of these two pairs of electrodes on the sphere was symmetrical in all the investigated scenarios, considering that the change of $\phi$ and $\theta$ was the same for all two pairs of electrodes Fig. 1.B. The stimulation area was evaluated using the $xy$ plane, and its depth was denoted by the $xz$ plane Fig. 1.C.

Our simulation analyzed how electrode placement and target location affect temporally interfering stimulation, establishing a relationship between electrode distance and stimulus depth.

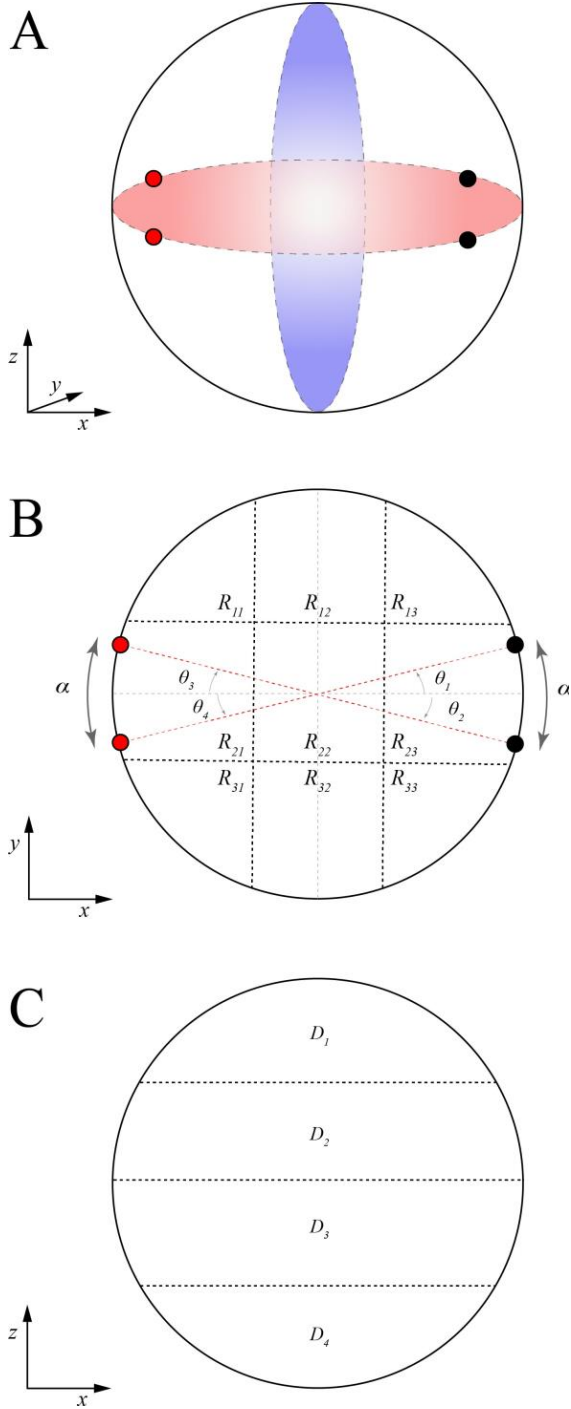

Fig. 1. A. Position of the electrode on the sphere (Note that the *xy* and *xz* planes are shown in red and blue, respectively), B. Electrodes $\theta$ and $\alpha$ are displayed in nine regions of the *xy* plane, C. The depth of the stimulation region is determined by utilizing the segmentation of the *xz* plane.

## IV. RESULTS

Our research delves into three scenarios in neural stimulation and brain modulation, each unraveling the intricate interplay of parameters $\phi$, $\theta$, and current ratio in brain neurostimulation applications. In the following, we will examine these scenarios and the effect of each parameter in the control of the stimulation area.

*a) First Scenario:* This scenario focuses on the effect of the parameter $\phi$ on the stimulation pattern as shown in Fig. 2. The right pair has a value of $-20°$ and $20°$, while the left pair has a value of $-160°$ and $160°$. The current ratio of electrode pairs is also constant. Intriguingly, our observations revealed that as we decreased the $\phi$ angle from $90°$ to $60°$ and subsequently to $30°$, the stimulation pattern transitioned from a more diffuse distribution to a highly focused one. This shift in stimulation concentration is paramount for neurostimulation and deep brain stimulation applications. Furthermore, the correlation between $\phi$ and the depth of stimulation is quite evident. When the $\phi$ was $90°$, the electrical current reached deeper into the targeted region. However, as we reduced $\phi$ to $60°$ and $30°$, the depth of stimulation progressively decreased. This phenomenon can have significant implications for medical interventions, as it allows for more precise targeting of neural structures with minimal impact on surrounding tissues.

*b) Second Scenario:* In this scenario, the fixed electrode positions, $\phi$, and $\theta$, are constant with $\phi$ for all electrodes at $70°$. $\theta$ for the right pairs is $10°$ and $-10°$, and $\theta$ for the left pairs is $-170°$ and $170°$. As we transition from the state ($I_L/I_R = 1$) to the subsequent state where ($I_L/I_R > 1$), we observe that the stimulation pattern shifts from center to right. Conversely, in another state where $0 < I_L/I_R < 1$, the stimulation region gravitates towards the left electrode pair, emphasizing that the stimulation's lateral positioning highly depends on the current ratio between the electrode pairs. These findings underscore the precise control achieved in neuromodulation by manipulating this current ratio Fig. 3. Importantly, while the lateral position of stimulation varies across these states, the stimulation depth remains a constant factor. This scenario suggests that by modulating the current distribution between electrode pairs, we can fine-tune the spatial focus of neural stimulation without altering the penetration depth. Such insights hold immense potential for designing more targeted and effective therapies in neuromodulation. Further research in this area is poised to unlock new avenues for personalized treatment approaches in neurological disorders.

*c) Third Scenario:* Now, we are exploring the impact of altering $\theta$ in a constant $\phi$ and electrode pairs ratio in the stimulation region and depth. When we adjust $\theta$, we inevitably influence $\alpha$. $\theta$ is the angle of each electrode with respect to the *x*-axis, and $\alpha$ is the angle between two electrode pairs ($\alpha = \theta_1 + \theta_2 = \theta_3 + \theta_4$), which plays a crucial role in determining the direction of stimulation. Our experiments considered $\alpha$ modes: $20°$, $60°$, and $100°$. In the first mode with $\alpha$ at $20°$, we observed a stimulation pattern that was centered and relatively balanced. However, as $\alpha$ increased to $60°$ in the second mode, the focal point of stimulation gradually shifted away from the center, indicating a directional bias in neural activation. The most intriguing observation came in the third mode, where $\alpha$ reached $100°$. In this configuration Fig. 4, stimulation at the center completely disappeared, and the pattern became exclusively located above and below the spherical electrode array. Remarkably, throughout all three modes, the stimulation depth remained consistent, emphasizing that variations in $\alpha$ primarily affect the spatial distribution of neural stimulation without compromising its penetration depth.

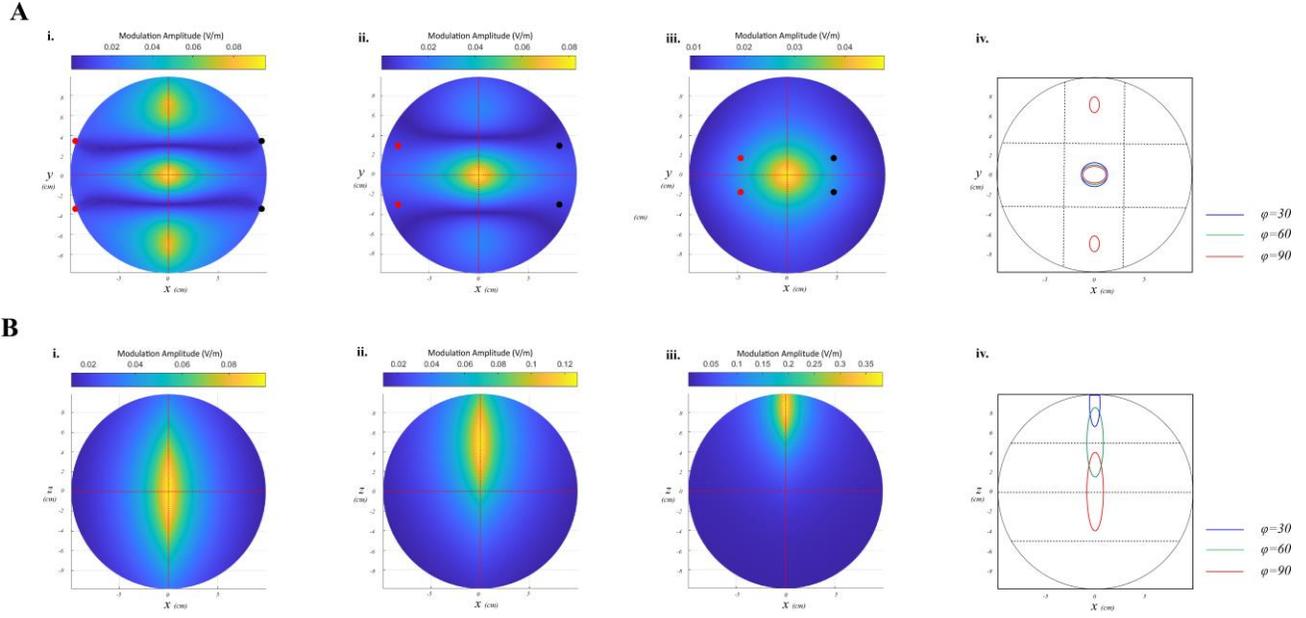

Fig. 2. A. Stimulation region in constant $\theta$ mode and currents ratio equals one. i. $\phi = 90°$ ii. $\phi = 60°$ iii. $\phi = 30°$ iv. The locus of the stimulation regions in a congruous state for these three scenarios. B. Stimulation depth in constant $\theta$ mode and current ratio equals one i. $\phi = 90°$ ii. $\phi = 60°$ iii. $\phi = 30°$ iv. The locus of the stimulation depth in a congruous state for these three scenarios.

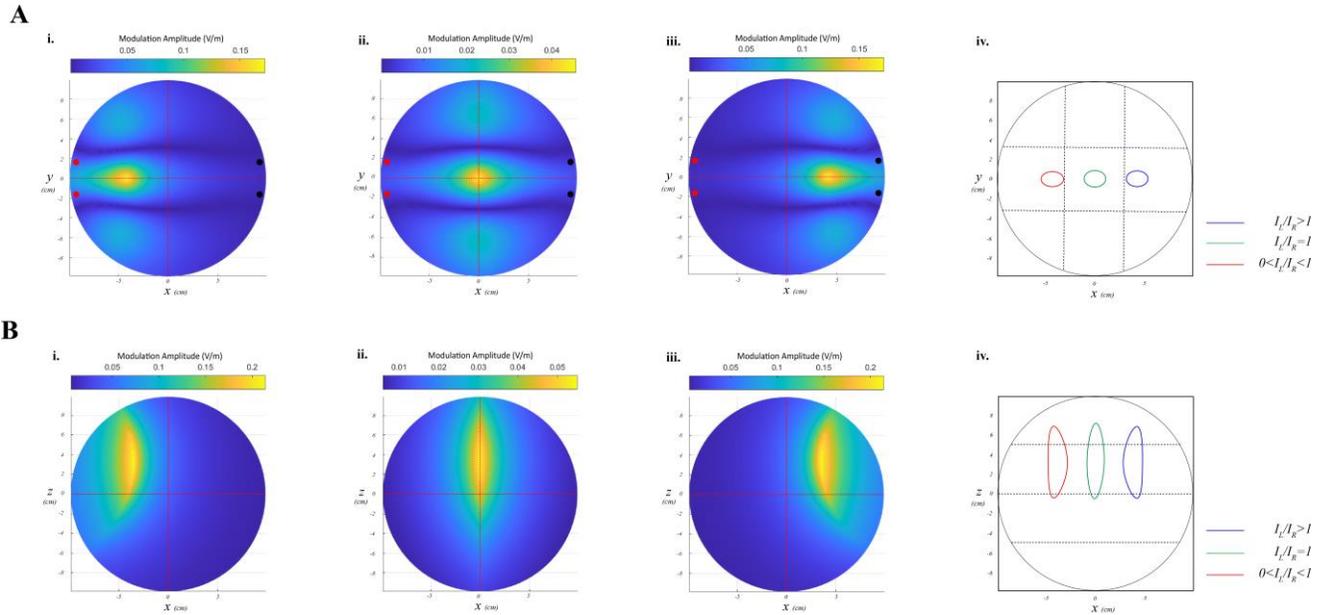

Fig. 3. Stimulation region for constant $\theta$ and $\phi$. i. $0 < I_L/I_R < 1$, ii. $I_L/I_R = 1$, iii. $I_L/I_R > 1$, iv. The locus of the stimulation region in a congruous state for these three scenarios B. Stimulation depth in constant of $\theta$ and $\phi$ i. $0 < I_L/I_R < 1$, ii. $I_L/I_R = 1$, iii. $I_L/I_R > 1$, iv. The locus of the stimulation depth in a congruous state for these three scenarios.

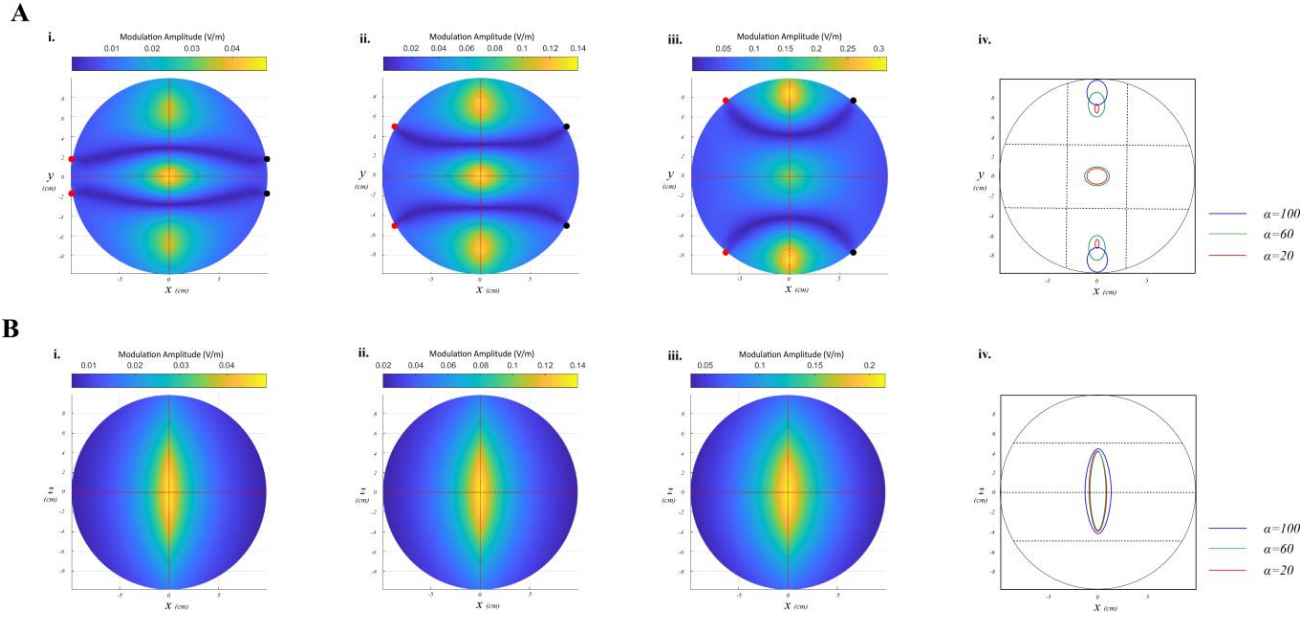

Fig. 4. A) Stimulation region in constant of $\phi$ and current ratio equals one i. $\alpha = 20°$ ii. $\alpha = 60°$ iii. $\alpha = 100°$ iv. The locus of the stimulation region in a congruous state for this three scenarios. B) Stimulation depth in a constant of $\phi$ and current ratio equals one i. $\alpha = 20°$ ii. $\alpha = 60°$ iii. $\alpha = 100°$ iv. The locus of the stimulation depth in a congruous state for these three scenarios.

TABLE I
GUIDELINE FOR STIMULATION PATTERNS IN NINE REGIONS OF THE $xy$ PLANE BY $xz$ PLANE SEGMENTATION TO DETERMINE THE DEPTH OF THE STIMULATION REGION STUDY ON FIVE DIFFERENT FEATURES

| Regions | Electrode Position | Exact Regions | Currents Ratio |
|---|---|---|---|
| $R_{11}, R_{12}, R_{13}$ | To stimulate regions $R_{11}$, $R_{12}$, and $R_{13}$, the electrode placement remains consistent throughout. The right pair must be set at angles ranging from $0°$ to $45°$ and $90°$ to $45°$, while the left pair must be positioned at angles between $90°$ to $135°$ and $180°$ to $135°$. | $R_{11}$ | To stimulate region $R_{11}$, the current of the right electrodes must be more than the left ones, and this ratio is $0 < I_L/I_R < 1$. |
| | | $R_{12}$ | Region $R_{12}$ is the center of this row, and the current ratio for this region must be equal to one, $I_L/I_R = 1$. |
| | | $R_{13}$ | Region $R_{13}$ is located on the right, and to stimulate this part, the left current must be greater than the right current, and the ratio is $I_L/I_R > 1$. |
| $R_{21}, R_{22}, R_{23}$ | In order to stimulate regions $R_{21}$, $R_{22}$, and $R_{23}$, the electrodes must be positioned consistently with the right pair set at angles ranging from $0°$ to $90°$ and $0°$ to $-90°$, while the left pair will be placed at angles between $90°$ to $180°$ and $-90°$ to $-180°$. | $R_{21}$ | To activate region $R_{21}$, the right electrode's current should exceed that of the left electrode, expressed as $0 < I_L/I_R < 1$. |
| | | $R_{22}$ | Region $R_{22}$ is situated at the row's midpoint, where the current ratio equals one, $I_L/I_R = 1$. |
| | | $R_{23}$ | In the case of region $R_{23}$, located on the right, the left electrode's current must surpass that of the right electrode, with a ratio greater than 1, represented as $I_L/I_R > 1$, to stimulate this area. |
| $R_{31}, R_{32}, R_{33}$ | For the stimulation of regions $R_{31}$ to $R_{33}$, the electrodes must be placed in a fixed position. The right pair will be angled between $0°$ to $-45°$ and $-90°$ to $-45°$, while the left pair will be positioned between $-90°$ to $-135°$ and $-180°$ to $-135°$. | $R_{31}$ | To trigger activity in region $R_{31}$, the current delivered through the right electrodes needs to be greater than that applied through the left electrodes, and this ratio should fall between 0 and 1, specifically $0 < I_L/I_R < 1$. |
| | | $R_{32}$ | Region $R_{32}$ occupies the central position in this row, where the current ratio remains at one, $I_L/I_R = 1$. |
| | | $R_{33}$ | Region $R_{33}$, located on the right side, requires a higher current through the left electrodes compared to the right electrodes, with a ratio exceeding 1, expressed as $I_L/I_R > 1$, to stimulate this particular region. |

## V. Discussion

Our study focused on noninvasive brain stimulation and explored three different scenarios. We analyzed the complex relationship between angular parameters and current ratios through these scenarios. We obtained graphical results by covering the surface of the model sphere and examining the stimulated region and depth. These findings enabled us to create a list of guidelines presented in Table I. To precisely identify the stimulation within each row in Tabe I, adjusting the current ratio between the left and right electrode pairs is essential. The $I_L/I_R$ ratio should be between zero and one for stimulating the left regions. For central region stimulation, the $I_L/I_R$ ratio should be strictly 1, while the $I_L/I_R$ ratio should be more than 1 for stimulating the right regions. The $xz$ plane (Fig. 1) indicates stimulation depth and was divided into four segments. The segments $D_1$, $D_2$, and $D_3$, must be selected to stimulate the upper, middle, and deep parts, respectively. Segment $D_4$ denotes an off-limit area due to limitations in electrode placement in the lower part of the skull, but the other three layers can be stimulated. We considered the range of angles in Table II to stimulate these segments. Hence, we can effectively assess and classify stimulation patterns in the segments of the $xz$ plane. The suggested segmentation allows us to target precise areas for electrode placement and stimulation within the brain. We can optimize therapeutic interventions and improve patient outcomes by categorizing these patterns. We generated Table II using this segmentation to establish classified patterns. The table was created to coordinate each specific scenario and serves as a valuable tool for clinicians to guide them in selecting the most appropriate stimulation strategy for each unique scenario. This enhances the overall effectiveness of the procedure.

## VI. Conclusion

We studied noninvasive brain stimulation, examining the changes in angular parameters and current ratios based on stimulation region and depth. Our findings have led us to establish a set of guidelines that can aid in the fine-tuning of noninvasive brain stimulation techniques. We believe that our study can have far-reaching implications in the field of neuromodulation, as it can help develop personalized and targeted treatment methods for neurological disorders. By understanding the influence of various factors on neural stimulation patterns, we hope that researchers can create more effective therapies, thus improving the quality of life for individuals with neurological ailments. However, our study has some practical limitations, as we only used a simple sphere model and considered only two pairs of electrodes. Therefore, the results obtained are not recommended for direct use in clinical practice.

TABLE II
Guideline for stimulation depth based on $xz$ plane

| Depth | Angles of the electrodes |
|---|---|
| $D_1$ | To stimulate the surface part of the cerebral cortex, $\phi$ must be within the range of $0°$ to $45°$. |
| $D_2$ | Stimulation of the subcortical region or middle depth of the brain, $\phi$ of electrodes must be between $45°$ and $90°$. |
| $D_3$ | For the target deep part of the brain, $\phi$ must be between $90°$ and $135°$. |
| $D_4$ | $\phi$ between $135°$ to $180°$ is an off-limit area due to limitations in electrode placement in the lower part of the skull. |